\documentstyle[multicol,aps,twocolumn,epsf]{revtex} 
\begin{document}
\draft
%\title{Predictability in the 2D inverse energy cascade}
\title{Predictability of the energy cascade in 2D turbulence}

\author{G. Boffetta and S. Musacchio}
\address{Dipartimento di Fisica Generale, Universit\`a di Torino,
         Via Pietro Giuria 1, 10125 Torino, Italy}
\address{and INFM Sezione di Torino Universit\`a, Italy}

\date{\today}

\maketitle

\begin{abstract}
The predictability problem in the inverse energy cascade of
two-dimensional turbulence is addressed by means of 
direct numerical simulations. 
The growth rate as a function of the error level is determined
by means of a finite size extension of the Lyapunov exponent.
For error within the inertial range, the
linear growth of the error energy, predicted by
dimensional argument, is verified with great accuracy.
Our numerical findings are in close agreement with the result 
of TFM closure approximation.
\end{abstract}


%%%%%%%%%%%%%%%%%%%%%%%%%%%%%%%%%%%%%%%%%%%%%%%%%%%%%%%%%%%%%%
Unpredictability is an essential property of turbulent flows.
Turbulence is characterized by a large number of degrees of
freedom interacting with a nonlinear dynamics. Thus turbulence
is chaotic (and hence unpredictable), but the standard approach
of dynamical system theory is not sufficient to characterize
predictability in turbulence. 

In fully developed turbulence, the maximum Lyapunov exponent 
diverges with the Reynolds number thus being
very large for typical turbulent flows.
Nevertheless, a large value of the Lyapunov exponent do not imply
automatically shot time predictability. A familiar example is
the atmosphere dynamics: Convective motions in the atmosphere
make the small scale features unpredictable after one hour or 
less, but large scale dynamics can be predicted for several
days, as it is demonstrated by weather forecasting.
This effect, which can be called ``strong chaos with weak
butterfly effect'' arises in systems possessing many 
characteristic scales and times.
From this point of view, turbulence probably represents the 
example most extensively studied.

The first attempts to the study of predictability in turbulence
dates back to the pioneering work of Lorenz \cite{Lorenz69} and to
Kraichnan and Leith papers \cite{Leith71,LK72}.
On the basis of closure approximations, it was possible to obtain
quantitative predictions on the evolution of the error in different
turbulent situations, both in two and three dimensions. 

A more recent approach to the problem is based on dynamical 
system theory. One of the first results is the Ruelle prediction
on the scaling of the Lyapunov exponent with the Reynolds 
number \cite{Ruelle79}. Chaotic properties 
have been extensively investigated in simplified models of
turbulence, called Shell Models, with particular emphasis on
the relations with intermittency \cite{OY89,JPV91}.
Because predictability experiments in fully developed turbulence 
are numerically rather expensive, a similar study on direct 
numerical simulations of Navier--Stokes equations is still lacking.

In this letter we address the predictability problem for 
two--dimensional turbulence by means of high resolution
direct numerical simulations.
Turbulence is generated in the inverse cascade regime
where a robust energy cascade is observed \cite{BCV00}. 
The absence of intermittency corrections makes the problem
simpler than in the three--dimensional case: velocity
statistics (energy spectrum, structure functions) is found to be
in close agreement with self-similar theory \`a la Kolmogorov.

%%%%%%%%%%%%%%%%%% FIGURE 
\narrowtext
\begin{figure}
\epsfxsize=8truecm
\epsfbox{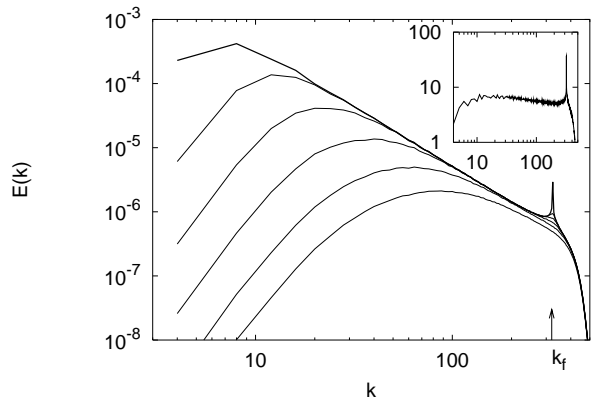}
\vspace{-0.2cm}
\caption{Stationary energy spectrum $E(k)$ (thick line) and 
error spectrum $E_{\Delta}(k,t)$ at time
$t=0.1, 0.2, 0.4, 0.8, 1.6$. $k_f=320$
is the forcing wavenumber. In the inset we plot the
compensated spectrum $\epsilon^{-2/3}k^{5/3}E(k)$.}
\label{fig:1}
\end{figure}
%%%%%%%%%%%%%%%%%%%%%%%%%%

The model equation is the two--dimensional Navier--Stokes equation
written for the scalar vorticity 
$\omega({\bf r},t)=- \triangle \psi({\bf r},t)$
with generalized dissipation and linear friction
\begin{equation}
\partial_t \omega + J(\omega,\psi) =
(-1)^{p+1} \nu_{p} \triangle^{p} \omega
- \alpha \omega - f
\label{eq:1}
\end{equation}
where $J$ represent the Jacobian with the stream function $\psi$
from which the velocity is ${\bf v}=(\partial_y \psi,-\partial_x \psi)$.
$p$ is the order of the dissipation, $p=1$ for ordinary
dissipation, $p>1$ for hyperviscosity.
As it is customary in numerical simulations, we use hyperviscous 
dissipation ($p=8$) in order to extend the inertial range.
Although this can affect the small scale features of the 
vorticity field \cite{Borue93}, in our simulations dissipation
is not involved in the cascade and has simply the role of removing
enstrophy at small scales.
The friction term in (\ref{eq:1}) removes energy at large scales: 
it is necessary in 
order to avoid Bose--Einstein condensation on the gravest mode \cite{SY93}
and to obtain a stationary state.
Energy is injected into the system by a random forcing
$\delta$--correlated in time which is active on a shell of wavenumbers
around $k_{f}$ only. 
Numerical integration of (\ref{eq:1}) is performed
by a standard pseudo-spectral code fully dealiased with second--order
Adams--Bashforth time stepping on a doubly periodic square domain
with resolution $N=1024$.

Stationary turbulent flow is obtained after a very long simulation
starting from a zero vorticity initial field. At stationarity we observe
a wide inertial range with a well developed Kolmogorov 
energy spectrum $E(k) = C \epsilon^{2/3} k^{-5/3}$ (Figure~\ref{fig:1}).
Structure functions in physical space are found in agreement 
with the self-similar Kolmogorov theory \cite{BCV00}.

Starting from the stationary configuration, the predictability experiment
integrates two different realizations of the turbulent field
and looks at the evolution of the difference (or error) field 
defined for the velocity coordinates as
\begin{equation}
\delta {\bf u}({\bf r},t) = {1 \over \sqrt{2}} 
\left({\bf u}_1({\bf r},t)-{\bf u}_2({\bf r},t)\right)
\label{eq:2}
\end{equation}

From (\ref{eq:2}) one defines the error energy
and the error energy spectrum as \cite{LK72,Lesieur97}
\begin{equation}
E_{\Delta}(t) = \int_{0}^{\infty} E_{\Delta}(k,t) \, dk =
{1 \over 2} \int |\delta {\bf u}({\bf r},t)|^2 d^2 r
\label{eq:4}
\end{equation}

Normalization in (\ref{eq:2}) ensures that $E_{\Delta}(k,t) \to E(k)$
for uncorrelated fields (i.e. for $t \to \infty$).

\begin{figure}
\epsfxsize=8truecm
\epsfbox{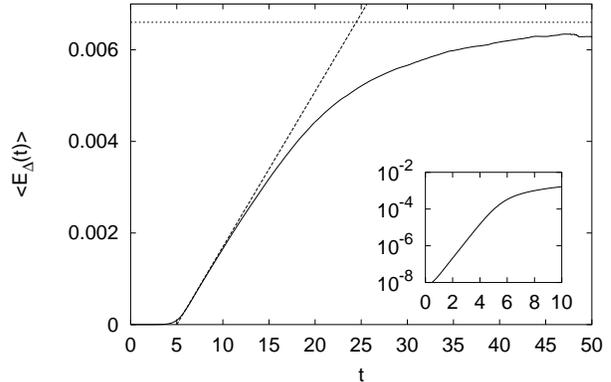}
%\vspace{-0.2cm}
\caption{Average error energy $\langle E_{\Delta}(t) \rangle$ growth. 
Dashed line represents closure prediction (\ref{eq:6}), 
dotted line is the saturation value $E$. 
The initial exponential growth is emphasized by the lin-log
plot in the inset.}
\label{fig:2}
\end{figure}

Assuming infinitesimal initial errors, the magnitude of 
the difference field starts to grow exponentially and
$E_{\Delta}(t) \simeq E_{\Delta}(0) exp(2 \lambda t)$
where $\lambda$ is the maximum Lyapunov exponent of the
system. The error growth is this stage is confined at the
faster scales in the inertial range, corresponding in our model 
to the scales close to the forcing wavenumber $k_{f}$ (see 
Figure~\ref{fig:1}). 
For finite errors, when $E_{\Delta}(k_{f},t)$ becomes
comparable with $E(k_{f})$, the exponential growth terminate
and an algebraic regime sets in. The dimensional prediction
proposed by Lorenz \cite{Lorenz69} assumes that the time
it takes for the error to induce a complete uncertainty at
wavenumber $k$ is proportional to the characteristic time 
at that scale, $t \simeq \tau(k)$. Within the Kolmogorov framework,
$\tau(k) \simeq \epsilon^{-1/3} k^{-2/3}$.
At larger scales the error is still very small in comparison with
the typical energy. We thus can write
\begin{equation}
E_{\Delta}(k',t=\tau(k)) = \left\{
\begin{array}{ll}
0 & \,\,\,\, \mbox{if $k'<k$} \\
E(k) & \,\,\,\, \mbox{if $k'>k$} 
\end{array}
\right.
\label{eq:5}
\end{equation}
By inserting (\ref{eq:5}) in (\ref{eq:4}), using the Kolmogorov 
spectrum for $E(k)$ and inverting the dimensional expression for
$\tau(k)$ one ends with the prediction \cite{Lorenz69,BJPV98}
\begin{equation}
E_{\Delta}(t) = G \epsilon t
\label{eq:6}
\end{equation}
The numerical constant $G$ in (\ref{eq:6}) can be obtained
only by repeating the argument more formally within a closure 
framework \cite{Leith71,LK72,ML86}.
In Figure~\ref{fig:2} we plot the time evolution of the error energy 
$\langle E_{\Delta}(t) \rangle$ obtained from direct 
numerical simulations averaged over $20$ realizations.
Both the exponential regime a small time and the 
linear regime (\ref{eq:6}) are visible. 

%%%%%%%%%%%%%%%%%% FIGURE 
\narrowtext
%%%%%%%%%%%%%%%%%%%%%%%%%%

The dimensional predictability argument given above can 
be rephrased in a language more close to dynamical systems by
introducing the Finite Size Lyapunov Exponent analysis.
FSLE is a generalization of the Lyapunov exponent to
finite size errors, which was recently proposed for the
analysis of systems with many characteristic scales
\cite{ABCPV96,ABCPV97,BGPV98}. In a nutshell
one computes the ``error doubling time'' 
$T_{r}(\delta)$, i.e. the time it
takes for an error of size $\delta=|\delta {\bf u}|$
to grow of a factor $r$ (for $r=2$ we have actually a 
doubling time).
The FSLE is defined in term of the average doubling time as
\begin{equation}
\lambda(\delta)={1\over \langle T_{r}(\delta) \rangle}
\ln r
\label{eq:7}
\end{equation}
It is easy to show that definition (\ref{eq:7}) reduces to the
standard Lyapunov exponent $\lambda$ in the infinitesimal
error limit $\delta \to 0$ \cite{ABCPV97}. 
For finite error, the FSLE measures the effective 
error growth rate at error size $\delta$.
Let us remark that taking averages at fixed time, as in (\ref{eq:6})
is not the same of averaging at fixed error size, as in (\ref{eq:7}).
This is particularly true in the case of intermittent systems,
in which strong fluctuations of the error in different realizations
can hide scaling laws like (\ref{eq:6}) \cite{ABCCV97}.
From a numerical point of view, the computation of $\lambda(\delta)$
is not more expensive than the computation of the Lyapunov 
exponent with a standard algorithm.

The same dimensional argument leading to (\ref{eq:6}), repeated
for $T_{r}(\delta)$, gives the prediction for the FSLE
in energy cascade inertial range
\begin{equation}
\lambda(\delta) = A \epsilon \delta^{-2}
\label{eq:8}
\end{equation}
where the constant $A$ is again not determined by dimensional arguments.

The scaling (\ref{eq:8}), which can be shown to be not affected by
possible intermittency corrections (as in 3D turbulence \cite{ABCPV96}), 
is valid within the inertial
range $u(k_{f}) < \delta < U$ where $u(k_f)$ represents the typical
velocity fluctuation at forcing wavenumber and $U\simeq \sqrt{2 E}$
is the large scale velocity. 

%%%%%%%%%%%%%%%%%% FIGURE 
\narrowtext
\begin{figure}
\epsfxsize=8truecm
\epsfbox{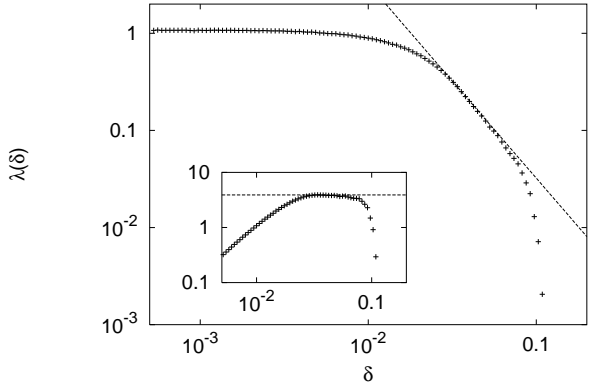}
%\vspace{-0.2cm}
\caption{Finite size Lyapunov exponent $\lambda(\delta)$ as a function
of velocity uncertainty $\delta$. The asymptotic constant value for
$\delta \to 0$ is the maximum Lyapunov exponent of the turbulent flow.
Dashed line represent prediction (\ref{eq:8}).
In the inset we show in the compensated plot 
$\lambda(\delta) \delta^{2}/\epsilon$. The line represent the 
fit to the constant $A \simeq 3.9$.}
\label{fig:3}
\end{figure}
%%%%%%%%%%%%%%%%%%%%%%%%%%

Figure~\ref{fig:3} shows the FSLE computed from our simulations.
For small errors $\delta < u(k_f)$ (corresponding to an error
spectrum $E_{\Delta}(k_f,t) << E(k_f)$) we observe the convergence of 
$\lambda(\delta)$ to the leading Lyapunov exponent. Its value is
essentially the inverse of the smallest characteristic time in the 
system and represents the growth rate of the most unstable features.
At larger $\delta > 10^{-2}$ we clearly see the transition to the
inertial range scaling (\ref{eq:8}). At further large 
$\delta \simeq U \simeq 0.1$,
$\lambda(\delta)$ falls down to zero in correspondence of error
saturation.

In order to emphasize scaling (\ref{eq:8}), in 
Figure~\ref{fig:3}
we also show the compensation of $\lambda(\delta)$ with 
$\epsilon \delta^{-2}$. Prediction (\ref{eq:8}) is verified with
very high accuracy which allows to determine the value of 
$A \simeq 3.9 \pm 0.1$. 
The constant $A$ relates the energy flux in the cascade to the 
rate of error growth. 
In absence of intermittency and with $r \simeq 1$ it is possible
to relate (\ref{eq:6}) to (\ref{eq:8}). In the present case
($r \simeq 1.12$) one obtains $G \simeq 4.1$.
It is interesting to observe that the numerical result is
very close to the old prediction obtained by
a Test Field Model closure \cite{LK72} which gives $G=4.19$.
At least from the point of view of predictability, two-dimensional
turbulence seems to be very well captured by low-order
closure scheme. 
As a consequence we can exclude, on the basis of our 
numerical findings, the existence of intermittency
effects in the inverse cascade of error.
This is a result which is probably of more 
general interest than the specific problem discussed in this letter.

\section*{Acknowledgments}
We are grateful to  A. Celani, M. Cencini and A. Vulpiani
for useful discussions.
Support from INFM ``PRA TURBO'', is gratefully
acknowledged.  Numerical simulations were partially performed 
at CINECA within the project "Fully developed two-dimensional
turbulence''.

%%%%%%%%%%%%%%%%%%%%%%%%%%%%%%%%%%%%%%%%%%%%%%%%%%%%%%%%%%%%%%%%%%%

\end{document}